\documentclass[11pt,a4paper]{article}

\usepackage{jheppub}
\usepackage{amsmath}
\usepackage{amsfonts}
\usepackage{bbm}
\usepackage{color}
\usepackage{microtype}
\usepackage{hyperref}
\usepackage{cleveref}

\makeatletter
\def\@fpheader{Prepared for submission to JCAP}
\makeatother

\newcommand{\sdg}{\sqrt{-g}}

\newcommand{\sdgb}{\sqrt{-g_\eff}}
\newcommand{\mn}{{\mu\nu}}

\newcommand{\calL}{\mathcal{L}_m}
\newcommand{\eff}{\mathrm{eff}}
\newcommand{\Mp}{M_\mathrm{Pl}}
\newcommand{\mc}{\mathcal{C}}

\makeatletter
\DeclareRobustCommand{\rcite}[1]{%
  \rcite@aux#1,\@nil{#1}%
}
\def\rcite@aux#1,#2\@nil#3{%
  \if\relax#2\relax
    Ref.~\cite{#3}%
  \else
    Refs.~\cite{#3}%
  \fi
}
\makeatother

\begin{document}

\title{Cosmological viability of massive gravity with generalized matter coupling}

\author[a]{Adam R. Solomon,}
\author[b]{Jonas Enander,}
\author[c,d]{Yashar Akrami,}
\author[e]{Tomi S. Koivisto,}
\author[d]{Frank K\"onnig,}
\author[b]{and Edvard M\"ortsell}
\affiliation[a]{DAMTP, Centre for Mathematical Sciences, University of Cambridge\\ Wilberforce Rd., Cambridge CB3 0WA, UK}
\affiliation[b]{Oskar Klein Center, Stockholm University,\\Albanova University Center\\ 106 91 Stockholm, Sweden}
\affiliation[c]{Institute of Theoretical Astrophysics, University of Oslo\\ P.O. Box 1029 Blindern, N-0315 Oslo, Norway}
\affiliation[d]{Institut f\"ur Theoretische Physik, Ruprecht-Karls-Universit\"at
Heidelberg\\ Philosophenweg 16, 69120 Heidelberg, Germany}
\affiliation[e]{Nordita, KTH Royal Institute of Technology and Stockholm University\\
       Roslagstullsbacken 23, SE-10691 Stockholm, Sweden}
\emailAdd{a.r.solomon@damtp.cam.ac.uk}
\emailAdd{enander@fysik.su.se}
\emailAdd{akrami@thphys.uni-heidelberg.de}
\emailAdd{tomi.koivisto@nordita.org}
\emailAdd{koennig@thphys.uni-heidelberg.de}
\emailAdd{edvard@fysik.su.se}

\abstract{There is a no-go theorem forbidding flat and closed FLRW solutions in massive gravity on a flat reference metric, while open solutions are unstable. Recently it was shown that this no-go theorem can be overcome if at least some matter couples to a hybrid metric composed of both the dynamical and the fixed reference metric. We show that this is not compatible with the standard description of cosmological sources in terms of effective perfect fluids, and the predictions of the theory become sensitive either to the detailed field-theoretical modelling of the matter content or to the presence of additional dark degrees of freedom. This is a serious practical complication. Furthermore, we demonstrate that viable cosmological background evolution with a perfect fluid appears to require the presence of fields with highly contrived properties. This could be improved if the equivalence principle is broken by coupling only some of the fields to the composite metric, but viable self-accelerating solutions due only to the massive graviton are difficult to obtain.}

\preprint{NORDITA-2014-107, HIP-2014-35/TH}

\maketitle

\section{Introduction}

Massive gravity has a long history \cite{Fierz:1939ix,Boulware:1973my,Vainshtein:1972sx,ArkaniHamed:2002sp,Creminelli:2005qk}, but only recently has the fully nonlinear, consistent theory of a massive graviton been constructed by de Rham, Gabadadze, and Tolley (dRGT) \cite{deRham:2010ik,deRham:2010kj,deRham:2011rn,deRham:2011qq,
Hassan:2011vm,Hassan:2011hr} (see \rcite{deRham:2014zqa} for a comprehensive review). However, this theory does not possess flat or closed Friedmann-Lema\^itre-Robertson-Walker (FLRW) cosmological solutions with a flat reference metric \cite{D'Amico:2011jj}, and the solutions which do exist, by choosing open curvature or a different reference metric, are unstable to the Higuchi ghost \cite{Higuchi:1986py} or other linear and nonlinear instabilities
\cite{Gumrukcuoglu:2011ew,Gumrukcuoglu:2011zh,Vakili:2012tm,DeFelice:2012mx,Fasiello:2012rw,DeFelice:2013awa}.

The search for viable cosmologies with a massive graviton has involved two routes. One is to extend dRGT by adding extra degrees of freedom. For example, these problems are at least partially cured in bigravity, where the second metric is given dynamics \cite{Hassan:2011tf,Hassan:2011zd,Hassan:2011ea,vonStrauss:2011mq,Akrami:2012vf,Akrami:2013pna,Konnig:2013gxa,Berg:2012kn,Solomon:2014dua,Konnig:2014xva,Lagos:2014lca,Enander:2015vja,Akrami:2015qga}. Other extensions of massive gravity, such as quasidilaton \cite{D'Amico:2012zv}, varying-mass \cite{D'Amico:2011jj,Huang:2012pe}, nonlocal \cite{Jaccard:2013gla,Foffa:2013vma,Dirian:2014ara}, and Lorentz-violating \cite{Comelli:2013paa,Comelli:2013tja} massive gravity, also seem to possess improved cosmological behavior. The other approach is to give up on homogeneity and isotropy. While FLRW solutions are important for their mathematical simplicity, which renders them easy both to compute and to compare to observations, the Universe could in principle have anisotropies which have such low amplitude, are so much larger than our horizon, or both, that we cannot readily observe them. Remarkably, these cosmologies not only exist in massive gravity but are locally (i.e., within the horizon) arbitrarily close to the standard FLRW case \cite{D'Amico:2011jj}. The general scenario of an FLRW metric with inhomogeneous St\"uckelberg fields has been derived in \rcite{Volkov:2012cf,Volkov:2012zb}. This includes, but is not limited to, the case in which the reference metric is still Minkowski space, but only has the canonical form $\eta_\mn = \operatorname{diag}(-1,1,1,1)$ in coordinates where $g_\mn$ is not of the FLRW form \cite{Gratia:2012wt}. The inhomogeneous and anisotropic solutions are reviewed thoroughly in \rcite{deRham:2014zqa}. See \rcite{DeFelice:2013bxa} for a review of cosmology in massive gravity and some of its extensions.

Recently a workaround that allows consistent flat FLRW solutions in the context of dRGT massive gravity --- i.e., the theory with only a single, massive graviton --- was discovered in \rcite{deRham:2014naa}. This solution is based on the fact that massive gravity contains a fixed reference metric, and matter can in principle couple to both metrics \cite{Akrami:2013ffa,Akrami:2014lja,Khosravi:2011zi}, although care must be taken to ensure this coupling does not reintroduce the exorcised ghost \cite{Yamashita:2014fga,deRham:2014naa}. In this scenario, matter is coupled to an effective or Jordan-frame metric given by
\begin{equation}
g_\mn^\eff \equiv \alpha^2g_\mn + 2\alpha\beta g_{\mu\alpha}X^\alpha_\nu + \beta^2\eta_\mn, \label{eq:geffdef}
\end{equation}
where $g_\mn$ is the dynamical metric, $\eta_\mn$ is the Minkowski reference metric, and $X^\mu_\nu\equiv(\sqrt{g^{-1}\eta})^\mu_\nu$. This effective metric was arrived at in the vielbein formulation by a complementary derivation in \rcite{Noller:2014sta}, and is claimed to be ghost-free at least within the effective theory's r\'egime of validity \cite{deRham:2014naa,deRham:2014fha,Hassan:2014gta}. In \rcite{deRham:2014naa}, it was shown that flat FLRW solutions exist when $\alpha, \beta\neq0$, and a worked example was presented in which one or more scalar fields couples to $g_\mn^\eff$. The matter coupling has since been studied in the context of bigravity in \rcite{Enander:2014xga,Schmidt-May:2014xla}, where it was shown to be consistent with observational data of the cosmic expansion.

In this paper we explore the basic properties of these newly-allowed massive cosmologies from an observationally-oriented standpoint. Unusually, the proof that FLRW cosmologies exist leans heavily on the choice of a fundamental field as the matter source coupled to the effective metric. In a standard late-Universe setup where matter is described by a perfect fluid with constant equation of state $w$ (or even more generally when $w$ only depends on the scale factor), this result does not hold, and FLRW solutions are constrained to be nondynamical, just as in standard dRGT. More generally, the pressure of at least one of the matter components coupled to $g_\mn^\eff$ must depend on something besides the scale factor --- such as the lapse or the time derivative of the scale factor --- for massive-gravity cosmologies to be consistent. This is why fields, which have kinetic terms where the lapse necessarily appears, are good candidates to obtain sensible cosmological solutions. Consequently the standard techniques of late-time cosmology cannot be applied to this theory. We emphasize this does not necessarily imply that cosmological solutions do not exist, but rather that we must either employ a more sophisticated description of the matter sector or include new degrees of freedom in order to obtain realistic models which can be reliably confronted with data.

Our focus here is on models with an extra, ``dark'' scalar degree of freedom coupled to $g_\mn^\eff$. While we do not aim to rule these out, we show that these solutions exhibit pathologies in the early- and late-time limits if all matter couples to the effective metric, and the scalar-field physics would need to be highly contrived to avoid these issues, although these pathologies are largely avoided if the equivalence principle is broken and only the new dark sector couples to $g^\eff_\mn$. Moreover, the reliance on a dark sector which may well be gravitationally subdominant and high-energy implies a violation of the decoupling principle, in which the low-energy expansion of the Universe should not be overly sensitive to high-energy physics.

During the completion of this paper, \rcite{Gumrukcuoglu:2014xba} appeared which studied the background cosmology of this theory with a scalar field coupled to the effective metric, and demonstrated its perturbative stability. We agree with their results wherever we overlap. Our emphasis differs, however, as we focus on the effects of the perfect fluids, particularly dust and radiation, expected to be gravitationally dominant in the late Universe.

The rest of this paper is organized as follows. In \cref{sec:cosmo-back-drgt} we derive and discuss the cosmological evolution equations in this theory. In \cref{sec:dyn-sol} we elucidate the conditions under which the no-go theorem is violated and dynamical cosmological solutions exist. We discuss in \cref{sec:einvsjord} some of the nonintuitive features of the Einstein-frame formulation of the theory, and how these are resolved in the Jordan-frame description. In \cref{sec:drgt-scalar} we study cosmologies containing only a scalar field, and generalize this to include a perfect fluid coupled to the effective metric in \cref{sec:scalandfluid}. In \cref{sec:mixedcouplings} we consider an alternative setup in which the scalar field couples to the effective metric while the perfect fluid couples to the dynamical metric. We discuss our results and conclude in \cref{sec:dc-drgt-summary}.

\section{Cosmological backgrounds}
\label{sec:cosmo-back-drgt}

If all matter fields couple to $g^\eff_\mn$, the theory is defined by the action
\begin{align}
  S &= - \frac{\Mp^2}{2}\int d^4x\sqrt{-g}R + m^2\Mp^2\int d^4x\sqrt{-g}\sum_{n=0}^{4}\beta_ne_n\left(X\right) \nonumber\\
  &\hphantom{{}=} + \int d^4x\sqrt{-g_\eff}\mathcal{L}_m\left(g_\eff, \Phi\right), \label{eq:action} 
\end{align}
where $e_n$ are the elementary symmetric polynomials of the eigenvalues of $X$, $\beta_n$ are dimensionless free parameters characterizing the strength of the different graviton interactions, and $\Phi$ represents the matter fields.
The Einstein equation for this theory was derived in \rcite{Schmidt-May:2014xla} and can be written in the form\footnote{Our convention is that indices on the Einstein tensor $G^\mn$ are raised with $g^\mn$.}
\begin{equation}
 (X^{-1})^{(\mu}{}_\alpha G^{\nu)\alpha} + m^2\displaystyle\sum_{n=0}^{3}(-1)^n\beta_ng^{\alpha\beta}(X^{-1})^{(\mu}{}_\alpha Y^{\nu)}_{(n)\beta} = \frac{\alpha}{\Mp^2}\det\left(\alpha+\beta X\right)\left(\alpha (X^{-1})^{(\mu}{}_\alpha T^{\nu)\alpha}+\beta T^{\mu\nu}\right), \label{eq:einstein}
\end{equation}
where the stress-energy tensor is defined in the usual way with respect to the effective metric,
\begin{equation}
T^\mn = \frac{2}{\sdgb}\frac{\delta\left[\sdgb\calL\left(g^\eff_\mn,\Phi\right)\right]}{\delta g^\eff_\mn},
\end{equation}
and the matrices $Y_{(n)}$ are given by
\begin{align}
  Y_{(0)} &\equiv \mathbbm{1}, \nonumber \\
  Y_{(1)} &\equiv X - \mathbbm{1}\left[X\right], \nonumber \\
  Y_{(2)} &\equiv X^2 - X\left[X\right] + \frac{1}{2}\mathbbm{1}\left(\left[X\right]^2 - \left[X^2\right]\right), \nonumber \\  
  Y_{(3)} &\equiv X^3 - X^2\left[X\right] + \frac{1}{2}X\left(\left[X\right]^2 - \left[X^2\right]\right) \nonumber \\
  &\hphantom{{}\equiv} - \frac{1}{6}\mathbbm{1}\left(\left[X\right]^ 3 - 3\left[X\right]\left[X^2\right] + 2\left[X^3\right]\right).
\end{align}
Notice that for diagonal metrics, including the FLRW metric, the symmetrization in the Einstein equation can be dropped and we can obtain a simpler version,
\begin{equation}
 G^\mn + m^2\displaystyle\sum_{n=0}^3(-1)^n\beta_ng^{\mu\alpha}Y^\nu_{(n)\alpha} = \frac{\alpha}{\Mp^2}\operatorname{det}(\alpha + \beta X)\left(\alpha T^\mn + \beta X^\mu_\alpha T^{\nu\alpha}\right).
\end{equation}

Let us assume a flat FLRW ansatz for $g_\mn$ of the form
\begin{equation}
 g_\mn dx^\mu dx^\nu = -N^2(t)dt^2 + a^2(t)\delta_{ij}dx^idx^j, \label{eq:FRW}
\end{equation}
and choose unitary gauge for the St\"uckelberg fields, $\eta_\mn=\operatorname{diag}(-1,1,1,1)$, so the effective metric is given by
\begin{equation}
 g^\eff_\mn dx^\mu dx^\nu = -N_\eff^2(t)dt^2 + a_\eff^2(t)\delta_{ij}dx^idx^j,
\end{equation}
where the effective lapse and scale factor are related to $N$ and $a$ by
\begin{equation}
 N_\eff = \alpha N + \beta, \qquad a_\eff = \alpha a + \beta. \label{eq:Naeff}
\end{equation}
We will define the Hubble rates for $g_\mn$ and $g_\mn^\eff$ by
\begin{equation}
 H \equiv \frac{\dot a}{aN},\qquad H_\eff \equiv \frac{\dot a_\eff}{a_\eff N_\eff}. \label{eq:Hdef}
\end{equation}
Notice that these are defined slightly differently than usual because of the inclusion of the lapse. This is because in diffeomorphism-invariant theories, such as general relativity, the lapse can be fixed by a gauge transformation and so corresponds to a choice of time coordinate. Indeed, in such a theory $\dot a/aN$ would simply be the Hubble rate defined in cosmic time (i.e., $\dot a/a$ with $N=1$). Because we do not have this freedom in massive gravity (once we have fixed the St\"uckelberg fields), we cannot freely choose a time coordinate in this way, and neither the lapse nor the time coordinate, $t$, has any physical meaning on its own. Instead these quantities will only appear through the combinations $Ndt$ and $N_\eff dt$. This motivates the Hubble rates we have defined in \cref{eq:Hdef}, which are simply $d\ln a/Ndt$ and $d\ln a_\eff/N_\eff dt$.

Now let us derive the cosmological equations of motion. The time component of \cref{eq:einstein} yields the Friedmann equation,
\begin{equation}
3H^2 = \frac{\alpha\rho}{\Mp^2}\frac{a_\eff^3}{a^3} + m^2\left(\beta_0 + \frac{3\beta_1}{a} + \frac{3\beta_2}{a^{2}} + \frac{\beta_3}{a^{3}}\right), \label{eq:fried}
\end{equation}
where $\rho \equiv -g^\eff_{00}T^{00}$ is the density of the matter source coupled to $g_\mn^\eff$.\footnote{If we have additional matter coupled to $g_\mn$, its density will enter the Friedmann equation (\ref{eq:fried}) in the standard way.} The spatial component of \cref{eq:einstein} gives us the acceleration equation,
\begin{equation}\label{eq:acc}
3H^{2} + \frac{2\dot{H}}{N}+ \frac{\alpha p}{\Mp^{2}}\frac{N_{\eff}a_{\eff}^{2}}{Na^{2}} =m^{2}\left[\beta_{0}+\beta_{1}\left(\frac{1}{N}+\frac{2}{a}\right)+\beta_{2}\left(\frac{2}{aN}+\frac{1}{a^{2}}\right)+\frac{\beta_{3}}{Na^{2}}\right],
\end{equation}
where $p\equiv(1/3)g^\eff_{ij}T^{ij}$ is the pressure. Notice that the double coupling leads to a time-dependent coefficient multiplying the density and pressure terms in \cref{eq:fried,eq:acc} and hence a varying gravitational constant for cosmological solutions. The Friedmann equation for the effective Hubble rate, $H_\eff$, can be determined from \cref{eq:fried} by the relation
\begin{equation}
H_\eff = \alpha\frac{Na}{N_\eff a_\eff}H, \label{eq:Heffdef}
\end{equation}
which follows from \cref{eq:Naeff}. Note that for practical purposes one could freely set $\alpha=1$ here by rescaling $g_\mn$, $\Mp$, and $\beta_n$; only the ratio $\beta/\alpha$ is physical \cite{Enander:2014xga}.

Matter is covariantly conserved with respect to $g^\eff_\mn$,
\begin{equation}
\nabla^\eff_\mu T^\mn=0, 
\end{equation}
from which we can obtain the usual energy conservation equation written in terms of the effective metric,
\begin{equation}\label{eq:continuity}
\dot\rho + 3\frac{\dot a_\eff}{a_\eff}\left(\rho + p\right) = 0.
\end{equation}
As in general relativity, this holds independently for each species of matter as long as we assume that interactions between species are negligible. Finally, we can take the divergence of the Einstein equation (\ref{eq:einstein}) with respect to $g_\mn$ and specialize to the FLRW background to find, after imposing stress-energy conservation, the ``Bianchi constraint,''
\begin{equation}
m^2M_\mathrm{Pl}^2a^2P(a) \dot a = \alpha\beta a_\eff^2p \dot a, \label{eq:bianchi}
\end{equation}
where we have defined
\begin{equation}
P(a) \equiv \beta_1 + \frac{2\beta_2}{a} + \frac{\beta_3}{a^2}.
\end{equation}
This can equivalently be derived using \cref{eq:fried,eq:acc,eq:continuity}, as well as by leaving the St\"uckelberg fields unfixed (recall that we have been working in unitary gauge from the start) and taking their equation of motion \cite{D'Amico:2011jj,deRham:2014naa}. The pressure, $p$, appearing in \cref{eq:bianchi} is the total pressure of the Universe, or, if different species couple to different metrics, the total pressure of all matter coupled to $g^\eff_\mn$.

\section{When do dynamical solutions exist?}
\label{sec:dyn-sol}

In the original, singly-coupled formulation of massive gravity, $\beta=0$ and so the right-hand side of \cref{eq:bianchi} vanishes, with the result that $a$ is constrained to be constant. This is nothing other than the no-go theorem on flat FLRW solutions in massive gravity. A nondynamical cosmology is, of course, still a solution when $\alpha$ and $\beta$ are nonzero, in which case the values of $a$ and $N$ are determined from \cref{eq:fried,eq:acc}. The question is now under which circumstances the theory also allows for dynamical $a$. 

To begin with, let us follow the standard techniques of cosmology by modeling the matter as a perfect fluid with $p=w\rho$, where $w$ is either a constant or depends only on $a_\eff$. Assuming that $\dot{a}\neq0$, \cref{eq:bianchi} becomes
\begin{equation}
m^2M_\mathrm{Pl}^2a^2P(a) = \alpha\beta w a_\eff^2\rho. \label{eq:bianchipf}
\end{equation}
Notice that due to our equation of state, $\rho$ is a function only of $a$ (or, equivalently, $a_\eff$). To see this, consider \cref{eq:continuity} in the form
\begin{equation}
 \frac{d\ln\rho}{d\ln a_\eff} + 3\left[1+w(a_\eff)\right] = 0.
\end{equation}
Integrating this will clearly yield $\rho = \rho(a_\eff)$. Unless the left-hand side of \cref{eq:bianchipf} has exactly the same functional form for $a_\eff$ as the right hand side (which is, e.g., the case when $w=-1/3$ and $\beta_2=\beta_3=0$), this equation is not consistent with a time-varying $a$. The theory does therefore not give viable cosmologies where all matter coupled to $g^\eff_\mn$ is described with an equation of state $p=w\rho$ if $w$ is constant or depends only on the scale factor, as is the case with, e.g., a standard perfect fluid.

This conclusion is avoided if the pressure also depends on the lapse. In this case, \cref{eq:bianchi} becomes a constraint on the lapse, unlocking dynamical solutions.\footnote{Another possibility is that the pressure depends on $\dot a$. Given the functional form of this dependence, the effective Hubble parameter in terms of $a_\eff$ can then be determined by combining \cref{eq:Heffdef,eq:bianchi,eq:FEgen}. We do not discuss this case any further.} The most obvious way to obtain a lapse-dependent pressure is to source the Einstein equations with a fundamental field rather than an effective fluid. This was exploited by \rcite{deRham:2014naa} to find dynamical cosmologies with a scalar field coupled to $g^\eff_\mn$. We discuss this case in more detail below. Therefore, while physical dust-dominated solutions may exist, we must either include additional degrees of freedom or treat the dust in terms of fundamental fields.\footnote{We note, however, that if the pressure of the dust is truly taken to be vanishing on large scales, then it would seem from the Bianchi constraint (\ref{eq:bianchi}) that the no-go theorem is still a problem.} The standard methods of late-time cosmology cannot be applied to doubly-coupled massive gravity.

\section{Einstein frame vs. Jordan frame}
\label{sec:einvsjord}

Before examining the cosmological solutions in detail, it behooves us to further clarify the somewhat unusual differences between this theory's Einstein and Jordan frames. If all matter couples to the effective metric, then, as we show below, the Friedmann equation in the Einstein frame is completely independent of the matter content of the Universe (up to an integration constant which behaves like pressureless dust). In the Einstein-frame description, matter components with nonzero pressure affect the cosmological dynamics not through the Hubble rate but rather through the lapse, $N$. Because the lapse is involved in the transformation from the Einstein-frame $H$ to the Jordan-frame $H_\eff$, cf. \cref{eq:Heffdef}, the Jordan-frame Friedmann equation (corresponding to the observable Hubble rate) does depend on matter.

We proceed to demonstrate this explicitly. Regardless of the functional form of $p$, and whether or not it depends on the lapse, as long as $\dot a\neq 0$ the pressure is constrained by \cref{eq:bianchi} to have an implicit dependence on $a$ given by
\begin{equation}
p(a) = \frac{m^2M_\mathrm{Pl}^2a^2P(a)}{\alpha\beta a_\eff^2}. \label{eq:pa}
\end{equation}
The continuity equation (\ref{eq:continuity}) can then be integrated to obtain
\begin{equation}
\rho(a) = \frac{C}{a_\eff^3} - \frac{3m^2M_\mathrm{Pl}^2}{\beta a_\eff^3}\left(\frac{\beta_1}{3} a^3+\beta_2 a^2 + \beta_3 a \right), \label{eq:rhoa}
\end{equation}
where $C$ is a constant of integration. Inserting this into \cref{eq:fried}, we find a generic form for the Einstein-frame Friedmann equation,
\begin{equation}
3H^2=m^2 \left(c_0+3 \frac{c_1}{a}+ 3\frac{c_2}{a^2}+\frac{c_3}{a^3}\right),
\label{eq:FEgen}
\end{equation}
where we have defined the coefficients
\begin{align}
c_0 &\equiv  \beta _0 - \frac{\alpha}{\beta}  \beta_1, \nonumber \\
c_1 &\equiv \beta _1 - \frac{\alpha}{\beta}  \beta_2, \nonumber \\
c_2 &\equiv \beta _2 - \frac{\alpha}{\beta}  \beta_3, \nonumber \\
c_3 &\equiv \beta _3 + \frac{\alpha C}{m^2\Mp^2}. \label{eq:cidef}
\end{align}
Notice that the functional forms of $p(a)$, $\rho (a)$, and $H^2(a)$ are completely independent of the energy content of the Universe, except for an integration constant scaling like pressureless matter. It is interesting to note that in the vacuum energy case (see \rcite{Enander:2014xga}) with $\beta_n = (\alpha/\beta)\beta_{n+1}$, all of the $c_i$ coefficients apart from $c_3$ vanish. Therefore if the metric interactions took the form of a cosmological constant for $g_\mn^\eff$, then the Einstein-frame Friedmann equation would scale as $a^{-3}$.

We emphasize that the dependence of the Einstein-frame quantities solely on $a$ and the mass terms is interesting and is certainly unusual, but it does not mean that matter does not affect the cosmological dynamics; as discussed above, if all matter couples to $g_\mn^\eff$, then the observable Hubble rate is $H_\eff$, and this \emph{does} depend on the matter content. If not all matter were coupled to $g^\eff_\mn$ --- for example, if the standard model fields were coupled to $g_\mn$ (as it is argued they should in \rcite{deRham:2014naa}) --- then the expression (\ref{eq:rhoa}) for $\rho(a)$ would only apply to the total density of the matter coupling to the effective metric, while the density of the fields coupled to the dynamical metric would appear in the Friedmann equation for $H$ in the usual way.

\section{Massive cosmologies with a scalar field}
\label{sec:drgt-scalar}

Let us turn to the properties of cosmological solutions. Recall that if we include matter whose pressure does not only depend on the scale factor, $a_\eff$, then the Bianchi constraint (\ref{eq:bianchi}) may not rule out dynamical cosmological solutions. For a pressure that also depends on the lapse, \cref{eq:bianchi,eq:FEgen} determine $N$ and $H$, respectively. These can be used in turn to derive the Jordan-frame Friedmann equation. Because the lapse enters into the frame transformation (\ref{eq:Heffdef}), the Jordan frame can be sensitive to matter even though the Einstein frame is not. The lapse thus plays an important and novel role in massive gravity compared to general relativity.

As discussed above, lapse-dependent pressures are not difficult to obtain: they enter whenever we consider a fundamental field with a kinetic term. Consider a universe dominated by a scalar field, $\chi$, with a canonical kinetic term and an arbitrary potential.\footnote{We note here that, for illustrative purposes, all of our discussions of a scalar field will assume that it is canonical. The more general $P(X)$ case is discussed in some detail in \rcite{deRham:2014naa}.} Its stress-energy tensor is given by
\begin{equation}
 T^\mn = \nabla_\eff^\mu\chi\nabla_\eff^\nu\chi - \left(\frac 1 2 \nabla_\alpha\chi\nabla^\alpha_\eff\chi + V(\chi)\right)g_\eff^\mn,
\end{equation}
where $\nabla_\eff^\mu \equiv g_\eff^\mn \nabla^\eff_\nu$ and $V(\chi)$ is the potential. The density and pressure associated to $\chi$ are
\begin{equation}
\rho_\chi = \frac{\dot\chi^2}{2N_\eff^2}+V(\chi), \qquad p_\chi = \frac{\dot\chi^2}{2N_\eff^2}-V(\chi). \label{eq:scalarrhop}
\end{equation}
The constraint (\ref{eq:bianchi}) now has a new ingredient; the lapse, $N_\eff$, which appears through the scalar field pressure.\footnote{The $\alpha_2$ theory studied in \rcite{deRham:2014naa} can be obtained by setting $\beta_0=3$, $\beta_1=-3/2$, $\beta_2=1/2$, and $\beta_3=0$ \cite{Hassan:2011vm}. With this parameter choice, the Bianchi constraint (\ref{eq:bianchi}) reproduces eq. (5.8) of \rcite{deRham:2014naa}. \label{foot:alpha2}}

One can then use the Bianchi identity to solve for the lapse and substitute it into the Friedmann equation to obtain an equation for the cosmological dynamics that does not involve the lapse \cite{deRham:2014naa}. A simple way to substitute out the lapse is to use the relation, following straightforwardly from \cref{eq:bianchi},
\begin{equation}
\frac{\dot\chi^2}{2N_\eff^2} = V(\chi) + \frac{m^2M_\mathrm{Pl}^2a^2P(a)}{\alpha\beta a_\eff^2}, \label{eq:K}
\end{equation}
as the lapse only appears in the Einstein-frame Friedmann equation through $\dot\chi^2/2N_\eff^2$. Note however that we can also use \cref{eq:K} to solve for the potential, $V(\chi)$, and write the Einstein-frame Friedmann equation in a form that does not involve the potential. Of course, if we were to additionally integrate the continuity equation as discussed above, then the Einstein-frame Friedmann equation would take the form of \cref{eq:FEgen} which contains neither the kinetic nor the potential term.

Using \cref{eq:pa,eq:rhoa} we can find expressions for the kinetic and potential energies purely in terms of $a$,
\begin{align}
K(a) &= \frac{m^2M_\mathrm{Pl}^2a^3}{2\alpha a_\eff^3}\left(\frac{c_1}{a} + 2\frac{c_2}{a^2}+\frac{c_3}{a^3}\right), \label{eq:Ka}\\
V(a) &= -\frac{m^2M_\mathrm{Pl}^2a^3}{2\alpha a_\eff^3}\left(2d_0+\frac{d_1}{a} + 2\frac{d_2}{a^2}+\frac{d_3}{a^3}\right), \label{eq:Va}
\end{align}
where $K \equiv \dot\chi^2/2N_\eff^2$, the $c_i$ are defined in \cref{eq:cidef}, and we have further defined
\begin{align}
d_0 &\equiv  \frac{\alpha}{\beta}  \beta_1, \nonumber \\
d_1 &\equiv \beta _1 + 5\frac{\alpha}{\beta}  \beta_2, \nonumber \\
d_2 &\equiv \beta _2 + 2\frac{\alpha}{\beta}  \beta_3, \nonumber \\
d_3 &\equiv \beta _3 - \frac{\alpha C}{m^2\Mp^2}.
\end{align}
The integration constant, $C$, appears when solving the continuity equation (\ref{eq:continuity}). The Friedmann equation is given by the generic \cref{eq:FEgen}. That is, we are left with the peculiar situation that the pressure, energy density, and Einstein-frame Friedmann equation are completely insensitive to the form of the scalar field potential. As discussed above, this lack of dependence on the details of the scalar field physics is illusory; the lapse does depend on $V(\chi)$ and $\dot\chi$, cf. \cref{eq:K}, and in turn the Jordan-frame expansion history depends on the lapse, cf. \cref{eq:Heffdef}.

Let us briefly remark on a pair of important exceptions. The no-go theorem forbidding dynamical $a$ still applies when there is a scalar field present if either the potential does not depend on the lapse (such as a flat potential) or the field is not rolling. Let us rewrite \cref{eq:continuity} (which is equivalent to the Klein-Gordon equation) as
\begin{equation}
\frac{d}{dt}\left(\frac{\dot\chi^2}{2N_\eff^2} + V(\chi)\right) + 3\frac{\dot a_\eff}{a_\eff}\frac{\dot\chi^2}{N_\eff^2} = 0. \label{eq:kg}
\end{equation}
If $V(\chi)$ is independent of $N_\eff$ then $\dot\chi^2/N_\eff^2$ cannot depend on $N_\eff$ and, by extension, neither can $p=\dot\chi^2/2N_\eff^2-V(\chi)$. In the specific case of $V(\chi)=\mathrm{const.}$ this is clearly true, and we find $\dot\chi^2/N_\eff^2 \propto a_\eff^{-6}$, so $p=p(a)$. Similarly, if the field is not rolling, $\dot\chi=0$, then it is clear from \cref{eq:scalarrhop} that $p$ loses its dependence on the lapse.

To conclude this section, when a scalar field is coupled to the effective metric, we avoid the no-go theorem and it is possible to have dynamical $a$, unless the potential does not depend on the lapse (including a constant potential) or the field is not rolling. This result agrees with and slightly generalizes that presented in \rcite{deRham:2014naa,Gumrukcuoglu:2014xba}. In a realistic scenario, however, we will have not only a scalar field but also matter components present. We now turn to that scenario.

\section{Adding a perfect fluid}
\label{sec:scalandfluid}

We have seen that the no-go theorem on FLRW solutions in dRGT massive gravity continues to hold in the doubly-coupled theory if the only matter coupled to the effective metric is a perfect fluid whose energy density and pressure depend only on the scale factor. This complicates the question of computing dust-dominated or radiation-dominated solutions in massive gravity. One solution might be to treat the dust in terms of fundamental fields. Another would be to add an extra degree of freedom such as a scalar field. Its role is to introduce a lapse-dependent term into the Bianchi constraint (\ref{eq:bianchi}) and thereby avoid the no-go theorem.

It is this possibility which we study in this section. In \cref{sec:drgt-scalar} we examined the scalar-only case. Let us now include other matter components, such as dust or radiation, also coupled minimally to $g^\eff_\mn$. We assume that the density and pressure of these matter components, $\rho_\mathrm{m}$ and $p_\mathrm{m}$, only depend on $a_\eff$.\footnote{As discussed above and in \rcite{deRham:2014naa}, in principle any dust or radiation is made of fundamental particles for which the stress-energy tensor does depend on the lapse. We introduce this effective-fluid description because it is the standard method of deriving cosmological solutions in nearly any gravitational theory and is thus an important tool for comparing to observations.} We can then write the total density and pressure as     
\begin{align}
\rho = K+V+\rho_\mathrm{m}, \nonumber\\
p = K-V+p_\mathrm{m},
\end{align}
so that
\begin{align}
K=\frac{\rho+p-(\rho_\mathrm{m}+p_\mathrm{m})}{2}, \nonumber \\
V=\frac{\rho-p-(\rho_\mathrm{m}-p_\mathrm{m})}{2}.
\label{eq:negK}
\end{align}
Note that \cref{eq:Ka,eq:Va} no longer hold, as they were derived in the absence of other matter, but \cref{eq:rhoa,eq:pa} are still valid and are crucial.

We would like to investigate the cosmological dynamics of this model. Rather than explicitly solving for the lapse and substituting it into the Friedmann equation for $H_\eff$, which leads to a very complicated result, we will take advantage of the known forms of $K(a_\eff)$ and $V(a_\eff)$, as well as the fact that $N_\eff$ only appears in $H_\eff$ and $K$ through the operator
\begin{equation}
\frac{d}{d\tau} = \frac{1}{N_\eff}\frac{d}{dt}.
\end{equation}
The physical Hubble rate is given by
\begin{equation}
H_\eff \equiv \frac{\dot a_\eff}{a_\eff N_\eff}=\frac{\alpha\dot a}{a_\eff N_\eff}.
\label{eq:heff2}
\end{equation}
Using the chain rule, we can write
\begin{equation}
\dot a=\frac{da}{dt}=\frac{da}{dV}\frac{dV}{d\chi}\frac{d\chi}{dt}=\frac{V'\dot\chi}{(dV/da)}, 
\end{equation}
where a prime denotes a derivative with respect to $\chi$. We also know that $\dot\chi=N_\eff\sqrt{2K}$, giving
\begin{equation}
\dot a=\frac{V'N_\eff\sqrt{2K}}{(dV/da)}, 
\end{equation}
which we can plug into \cref{eq:heff2} to obtain
\begin{equation}\label{eq:Heffscalar}
H^2_\eff =\frac{(V')^2 2K}{a^2_\eff (dV/da_\eff)^2}.
\end{equation}

This is the Friedmann equation for any universe with a scalar field rolling along a nonconstant potential. Every term in \cref{eq:Heffscalar} can be written purely in terms of $a_\eff$, allowing the full cosmological dynamics to be solved in principle. $K$ and $dV/da_\eff$ are given in terms of $a_\eff$ by \cref{eq:negK} [using \cref{eq:pa,eq:rhoa}]. $V'$ as a function of $a_\eff$ can be determined from the same equations once the form of $V(\chi)$ is specified. Note that while the lapse is not physically observable, its evolution in terms of $a$ can then be fixed by using \cref{eq:Heffdef} to find
\begin{equation}
\frac{N^2}{N_\eff^2}=2K\left(\frac{V'}{\alpha aH(dV/da_\eff)}\right)^2,
\end{equation}
where $H(a)$ is given by \cref{eq:FEgen}.

Assuming that the matter has a constant equation of state, we can use the known forms of $K(a)$ and $V(a)$ to find a relatively simple expression for the Friedmann equation up to $V'$,
\begin{equation}\label{eq:Hefffinal}
\left(\frac{H_\eff}{V'}\right)^2 = \frac{4\alpha^3\beta a_\eff^3\left(\mc_0 + \mc_1a_\eff + \mc_2 a_\eff^2 + \mc_\rho a_\eff^3\right)}{\left[3\mc_0 + 4\mc_1a_\eff + 5\mc_2 a_\eff^2 + 3(1-w)\mc_\rho a_\eff^3\right]^2},
\end{equation}
where for brevity we have defined
\begin{align}
\mc_0 &\equiv  \beta  \left[\alpha ^3 C+\beta ^2\beta _1+m^2\Mp^2 \left(3 \alpha  \left(\alpha  \beta _3-\beta  \beta_2\right) \right)\right], \nonumber \\
\mc_1 &\equiv -2 m^2\Mp^2 \left[\alpha  \left(\alpha  \beta _3-2 \beta  \beta _2\right)+ \beta^2\beta _1\right],  \nonumber \\
\mc_2 &\equiv m^2\Mp^2 \left(\beta  \beta _1-\alpha  \beta _2\right), \nonumber \\
\mc_\rho &\equiv -\alpha ^3 \beta  (1+w) \rho_\mathrm{m}.\label{eq:Hefffinalcoeff}
\end{align}
Notice that the right-hand side is a function of $a$ only.

The Friedmann equation~(\ref{eq:Hefffinal}) cannot be straightforwardly solved for generic choices of the potential, so we will make progress by examining past and future asymptotics, taking into account radiation ($w=1/3$) in the former and dust ($w=0$) in the latter. Before we do this, it is important to note that taking these asymptotics is not always simple, as we cannot necessarily assume that $a_\eff\to0$ at the beginning of the Universe or that $a_\eff\to\infty$ as $t\to\infty$. This means that, for example, our late-time analysis (in which $a_\eff$ is taken to infinity) will only be applicable for cosmologies in which the Universe expands \textit{ad infinitum}. Depending on the choice of scalar field potential, the Universe might end up, for example, recollapsing or approaching an asymptotic maximum value of $a_\eff$. A major aim of this section is to show the difficulties in obtaining standard cosmologies with a scalar field and perfect fluid both coupled to $g_\mn^\eff$; since a Universe which does not expand to infinity is highly nonstandard, we will find it sufficient to take $a_\eff\to\infty$ as the late-time limit in our search for viable cosmologies.\footnote{Of course, observations do not necessarily rule out the possibility of the scale factor not evolving to infinity, but it seems likely that making such a model agree with the data would require some serious contrivances.} We will see an example of when this limit may not be applicable.

Taking $a_\eff\to\infty$ in \cref{eq:negK}, we find
\begin{align}
\frac{\dot\chi^2}{2N_\eff^2} &\xrightarrow{a_\eff\to\infty} \frac{m^2 \Mp^2\left(\beta  \beta _1 - \alpha  \beta _2\right) }{2 \alpha ^3 \beta  a_\eff}, \\
V(\chi) &\xrightarrow{a_\eff\to\infty} -\frac{\beta _1 m^2 \Mp^2}{\alpha ^3 \beta }. \label{eq:latetimeV}
\end{align}
We see that the scalar field slows to a halt: $V(\chi)$ approaches a constant, while $d\chi/d\tau$, where $d\tau = N_\eff dt$ is the proper time, approaches zero. Notice that $V(\chi)$ is forced by the dynamics to approach a specific value, $V \to -\beta _1 m^2 \Mp^2/\alpha ^3 \beta$. \textit{A priori} there is no guarantee this value is within the range of $V(\chi)$, assuming the scalar field potential is not somehow set by gravitational physics. For example, a positive-definite potential like $V\sim\chi^2$ or $V\sim\chi^4$ would never be able to reach such a value, assuming $\alpha$, $\beta$, and $\beta_1$ are positive. Indeed, one can solve \cref{eq:negK} explicitly for $\chi(a_\eff)$ in such a case and find that, for large values of $a_\eff$, $\chi$ and $H_\eff$ are imaginary: there is a maximum value of $a_\eff$ at which $\chi^2$ and $H_\eff^2$ cross zero and become negative. Because such cosmologies are highly nonstandard and are unlikely to agree with data, we leave their study for future work.

Taking the large-$a_\eff$ limit of the Friedmann equation (\ref{eq:Hefffinal}), we obtain
\begin{equation}\label{eq:HefffinalLateTimes}
\left(\frac{H_\eff}{V'}\right)^2 \xrightarrow{a_\eff\to\infty} \frac{4\alpha^3\beta}{25\mc_2}a_\eff.
\end{equation}
Because $V(\chi)$ approaches a late-time value given by \cref{eq:latetimeV}, then assuming $V(\chi)$ is invertible, $\chi$ must also approach a constant $\chi_c$. This means that $V'=(dV/d\chi)|_{\chi=\chi_c}$ contributes a constant to \cref{eq:HefffinalLateTimes}. This is counter-intuitive; while the scalar field approaches a constant, $\dot\chi\to0$, $V'$ can and generically will approach a nonzero constant, which is just the slope of the potential evaluated at the asymptotic-future value of $\chi$, $\chi_c$. The Klein-Gordon equation (\ref{eq:kg}) is still satisfied because, as long as $V'$ does not go to zero, we can see from \cref{eq:HefffinalLateTimes} that $H_\eff\to\infty$ at late times. Therefore, the reason the scalar field slows down, in terms of the Klein-Gordon equation, is that the Hubble friction grows arbitrarily large, bringing the field to a halt even on a potential with a nonzero slope.\footnote{We thank the referee for helpful discussions on this point.} Unless the potential is contrived such that $V'\to0$ as $V\to-\beta_1m^2\Mp^2/\alpha^3\beta$, we see from \cref{eq:HefffinalLateTimes} that $H_\eff$ generically blows up, which is potentially disastrous behavior. This implies a violation of the null energy condition.

As we discuss below, if $V'$ goes to 0 then, depending on the speed at which it does so, $H_\eff$ may be better behaved. 

At early times, demanding the existence of a sensible radiation era leads to further problems. Assuming radiation couples to $g^\eff_\mn$, then $\rho_\mathrm{m} \sim a_\eff^{-4}$ with $p_\mathrm{m}=\rho_\mathrm{m}/3$. We have, cf. \cref{eq:negK}, that $2K = \rho + p - (\rho_\mathrm{m} + p_\mathrm{m})$, but, cf. \cref{eq:rhoa}, $\rho$ and $p$ do not have any terms scaling as steeply as $a_\eff^{-4}$. Therefore, in the presence of radiation, $\rho_\chi$ and $p_\chi$ pick up a \emph{negative} term going as $a_\eff^{-4}$ to exactly cancel out $\rho_\mathrm{m}$ and $p_\mathrm{m}$, leading to $K<0$ at sufficiently early times. From \cref{eq:Heffscalar} we see that this would lead to a negative $H_\eff^2$, and hence to an imaginary Hubble rate. Equivalently, we can take the early-time limit of \cref{eq:Hefffinal} to show, setting $\rho_\mathrm m = \rho_0 a_\eff^{-4}$,
\begin{equation}\label{eq:HefffinalEarlyTimes}
\left(\frac{H_\eff}{V'}\right)^2 \xrightarrow{a_\eff\to0} -\frac{3}{4\rho_0}a_\eff^4,
\end{equation}
so that again we see (for a real potential) $H_\eff$ becoming imaginary.

How could these conclusions be avoided? We can reproduce sensible behavior, but only if the potential is extremely contrived. At early times, we would need to arrange the scalar's dynamics so that $V'\to\infty$ ``before'' (i.e., at a later $a_\eff$ than) $K$ crosses zero.\footnote{The other obvious possibility, having $dV/da_\eff$ reach 0 before $K$ does, is impossible given the forms of $K(a)$ and $V(a)$.} We would then reach the initial singularity, $H_\eff\rightarrow\infty$, before the kinetic term turns negative.\footnote{This proposal has an interesting unexpected advantage: the Universe would begin at finite $a_\eff$, so a UV completion of gravity might not be needed to describe the Big Bang in the matter sector.} Moreover, we would need to tune the parameters of the theory so that $K=0$ happens at extremely early times, specifically before radiation domination. At intermediate times, $V'$ would need to scale in a particular way to [through \cref{eq:Hefffinal}] reproduce $H_\eff^2\sim a_\eff^{-4}$ and $H_\eff^2\sim a_\eff^{-3}$ during the radiation- and matter-dominated eras, respectively. Finally, in order to have $H_\eff\to\mathrm{const.}$ at late times, we see from \cref{eq:HefffinalLateTimes} that we would require $V'$ to decay as $a_\eff^{-1/2}$. We can construct such a potential going backwards by setting $H_\eff=H_{\Lambda\mathrm{CDM}}$ in \cref{eq:Hefffinal}, but there is no reason to expect such an artificial structure to arise from any fundamental theory. Even then we may still get pathological behavior: we can see from \cref{eq:Heffdef} that $N_\eff$ diverges if, at some point during the cosmic evolution, $H_\eff a_\eff = Ha$.

\section{Mixed matter couplings}
\label{sec:mixedcouplings}

Before concluding, we briefly discuss a slightly different formulation which avoids some of these problems. If we consider a scalar field and a perfect fluid, the avoidance of the no-go theorem on FLRW solutions only requires that the scalar field couple to $g^\eff_\mn$. In principle, all other matter could still couple to $g_\mn$. In fact, this is the theory that was studied in \rcite{deRham:2014naa}, where it was argued more generally that only a new dark sector should couple to $g^\eff_\mn$, while the standard model, as well as dark matter and dark energy, should couple to $g_\mn$. This theory violates the equivalence principle in the scalar sector, but is not \emph{a priori} excluded, and will turn out to have somewhat better cosmological behavior. Moreover, there is a compelling theoretical reason to consider such ``mixed'' couplings: matter loops would only generate a cosmological constant and would not destabilize the rest of the potential. This is because the vacuum energy associated to $g_\mn^\eff$ takes the form of the dRGT potential with all $\beta_n$ parameters nonzero, while the vacuum energy of matter coupled to $g_\mn$ only contributes to $\beta_0$ \cite{deRham:2013qqa}. We note that this problem may nevertheless persist with the dark fields that couple to $g_\mn^\eff$, unless their vacuum energy is somehow protected from loop corrections. We have seen in \cref{sec:drgt-scalar} that one simple possibility, using a massless field, does not seem to work because after integrating the Klein-Gordon equation, the pressure loses its dependence on the lapse.

Because the perfect fluid couples to $g_\mn$ and we derived the Bianchi constraint (\ref{eq:bianchi}) by taking the $g$-metric divergence of the Einstein equation, the constraint will now only contain $p_\chi$ rather than the total pressure, i.e.,
\begin{equation}
m^2M_\mathrm{Pl}^2a^2P(a) \dot a = \alpha\beta a_\eff^2p_\chi \dot a.
\end{equation}
This is the same constraint as in the scalar-only case discussed in \cref{sec:drgt-scalar}, so the scalar's kinetic and potential energies have the same forms, $K(a)$ and $V(a)$, as in \cref{eq:Ka,eq:Va}. The physical Hubble rate is now $H$, which after solving for the lapse is determined by the equation\footnote{Using the transformations to the $\alpha_2$ theory in \cref{foot:alpha2}, we recover eq. (5.9) of Ref.~\cite{deRham:2014naa}.}
\begin{equation}
3H^2=\frac{\rho_\mathrm{m}}{\Mp^2}+m^2 \left(c_0+3 \frac{c_1}{a}+ 3\frac{c_2}{a^2}+\frac{c_3}{a^3}\right),
\label{eq:friedconstrained}
\end{equation}
where the $c_i$ coefficients are defined in \cref{eq:cidef}. We emphasize that \cref{eq:friedconstrained} is completely generic when some matter couples to $g_\mn$ and some, possibly in a dark sector, couples to $g^\eff_\mn$. We have not assumed anything about the structure of the fields coupling to the effective metric, as we can derive \cref{eq:rhoa} for $\rho(a)$ and hence \cref{eq:friedconstrained} simply by using the Bianchi constraint to integrate the stress-energy conservation equation.

The cosmological behavior in this theory is fine. Because the scalar field does not have to respond to matter to maintain a particular form of $\rho(a)$ and $p(a)$, we no longer have pathological behavior in the early Universe, where there will be a standard $a^{-4}$ evolution. Moreover, as was pointed out in \rcite{Gumrukcuoglu:2014xba}, there is late-time acceleration: as $\rho_\mathrm{m}\to0$, $3H^2\to m^2(\beta_0-(\alpha/\beta)\beta_1)$, which, if positive, leads to an accelerating expansion.

However, these are not always \emph{self}-accelerating solutions. We will demand two conditions for self-acceleration: that the late-time acceleration not be driven by a cosmological constant, and that it not be driven by $V(\chi)$, as both of these could easily be accomplished without modifying gravity. In other words, we would like the effective cosmological constant at late times to arise predominantly from the massive graviton.

Let us start with the first criterion, the absence of a cosmological constant. One can write the dRGT interaction potential in terms of elementary symmetric polynomials of the eigenvalues of either $X \equiv \sqrt{g^{-1}f}$ or $\mathbb K \equiv \mathbb I - X$, with the strengths of the interaction terms denoted by $\beta_n$ in the first case and by $\alpha_n$ in the latter \cite{Hassan:2011vm,deRham:2014zqa}. What is notable is that $\alpha_0\neq\beta_0$: the cosmological constant is not the same in these two parametrizations. Terms proportional to $\sdg$ arise from the other interaction terms when transforming from one basis to the other. We have worked in terms of $\beta_n$ as it is mathematically simpler, but in massive gravity with a Minkowski reference metric, the presence of a Poincar\'e-invariant preferred metric allows for a more concrete definition of the cosmological constant in terms of $\alpha_n$.\footnote{We thank Claudia de Rham for helpful discussions on this point.} Consider expanding the metric as
\begin{equation}
g_\mn = \eta_\mn + 2h_\mn + h_{\mu\alpha}h_{\nu\beta}\eta^\mn.
\end{equation}
This expansion is useful because the metric is quadratic in $h_\mn$ but is fully nonlinear, i.e., we have not assumed that $h_\mn$ is small \cite{deRham:2014zqa}. In this language, the cosmological constant term, proportional to $\sdg$, can be eliminated by setting $\alpha_0 = \alpha_1 = 0$. Making this choice of parameter, and using the fact that $\alpha_n$ and $\beta_n$ are related by \cite{Hassan:2011vm}
\begin{equation}
 \beta_n = (4-n)!\displaystyle\sum_{i=n}^{4}\frac{(-1)^{i+n}}{(4-i)!(i-n)!}\alpha_i,
\end{equation}
we find the effective cosmological constant can be expressed in terms of $\alpha_{2,3,4}$ by
\begin{align}
\Lambda_\eff &= \frac{m^2}{3}\left(\beta_0-\frac\alpha\beta\beta_1\right) \nonumber\\
&= \frac{m^2}{3}\left[3 \alpha _2 \left(2+\frac\alpha\beta\right)-\alpha _3 \left(4+ 3\frac\alpha\beta\right)+\alpha _4 \left(1+\frac\alpha\beta\right)\right]. \label{eq:drgt-effcc}
\end{align}

Part of this constant comes from the fixed behavior of the scalar field potential.\footnote{Notice from \cref{eq:Ka} that, as in \cref{sec:scalandfluid}, the scalar field slows down to a halt at late times, so there is no contribution from the kinetic energy.} This piece is not difficult to single out: it consists exactly of the terms in \cref{eq:drgt-effcc} proportional to $\alpha/\beta$. Taking the late-time limit of \cref{eq:Va}, we can see that $V(\chi)$ asymptotes to
\begin{equation}
V(\chi) \xrightarrow{a_\eff\to\infty} -\frac{m^2\Mp^2\beta_1}{\alpha^3\beta}.
\end{equation}
Now consider the Friedmann equation in the form (\ref{eq:fried}) with, at late times, $\rho \to 0$. We can define a cosmological-constant-like piece solely due to the late-time behavior of $V$ given by
\begin{equation}
\Lambda_\chi \equiv \frac{\alpha V}{3\Mp^2}\left(\frac{a_\eff}{a}\right)^3 \xrightarrow{a_\eff\to\infty} \frac{m^2}{3}\frac\alpha\beta\left(3\alpha_2 - 3\alpha_3 + \alpha_4\right).
\end{equation}
Then \cref{eq:drgt-effcc} can simply be written in the form
\begin{equation}
\Lambda_\eff = \frac{m^2}{3}\left(6\alpha_2 - 4\alpha_3 + \alpha_4\right) + \Lambda_\chi = \frac{m^2}{3}\beta_0 + \Lambda_\chi,
\end{equation}
where in the last equality we mention that the residual term is nothing other than $m^2\beta_0/3$, which is simply a consistency check.

The modifications to gravity induced by the graviton mass therefore lead to a constant contribution to the Friedmann equations at late times, encapsulated in $m^2\beta_0/3$ (with $\alpha_0=\alpha_1=0$, so we do not identify this term with a cosmological constant). In a truly self-accelerating universe, this term should dominate $\Lambda_\chi$. If it did not, the acceleration would be partly caused by the scalar field's potential, and one could get the same end result in a much simpler way with, e.g., quintessence. For generic values of $\alpha_n$ and for $\beta\sim\mathcal{O}(1)$, both of these contributions are of a similar size and will usually have the same sign. To ensure self-accelerating solutions, one could, for example, tune the coefficients so that $3\alpha_2 - 3\alpha_3 + \alpha_4 = 0$ (the scalar field contributes nothing to $\Lambda_\eff$) or $3\alpha_2 - 3\alpha_3 + \alpha_4 < 0$ (the scalar field contributes negatively to $\Lambda_\eff$), or take $\beta\ll1$ (the scalar field contributes negligibly to $\Lambda_\eff$).

We end this section by briefly discussing the link between theory and observation in this particular model. One might worry that the predictivity of the theory is hurt by demanding that there be a new dark sector coupled to $g_\mn^\eff$. It is then natural to suspect that the task of confronting doubly-coupled massive gravity with observations is hopelessly dependent on the nature of this new dark sector, and the theory's parameters will consequently be more difficult to constrain. Yet we have seen in this section that that is not true: the Friedmann equation (\ref{eq:friedconstrained}) makes no reference to any details of the dark field or fields.\footnote{This is not the case when all matter couples to the effective metric, as observations would trace $g_\mn^\eff$, which \emph{is} sensitive to the nature of the dark sector, rather than $g_\mn$. We have, however, seen that the case where the standard model couples to $g_\mn$ is by far the best-behaved version.}  Recall from \cref{sec:einvsjord} that this is a consequence of the Bianchi constraint on the dark sector. We thus have the unusual result that the expansion history in the theory with a new dark sector, and nothing else, coupled to $g_\mn^\eff$ is completely insensitive to the nature of the dark fields.\footnote{See \rcite{deRham:2014naa} for a complementary derivation of this result.} There could be one scalar field or more, with any assortment of potentials and kinetic terms, and as long as they exist, and are subject to the technical conditions discussed above (such as having a nontrivial potential, if the kinetic term is canonical), then their contribution to the cosmological dynamics is given by the mass term in \cref{eq:friedconstrained}. This is good news for observers looking to perform geometrical tests of this theory. However, we are not aware of any reason that this lack of dependence on the details of the dark sector should extend beyond the simple background FLRW case. Even linear cosmological perturbations might be sensitive to the dark physics \cite{Gumrukcuoglu:2014xba}, which would present a challenge in comparing this theory to structure formation.

\section{Discussion and conclusions}
\label{sec:dc-drgt-summary}

One can extend dRGT massive gravity by allowing matter to couple to an effective metric constructed out of both the dynamical and the reference metrics. The no-go theorem ruling out flat or closed homogeneous and isotropic cosmologies in massive gravity \cite{D'Amico:2011jj} can be overcome when matter is ``doubly coupled'' in such a way \cite{deRham:2014naa,Gumrukcuoglu:2014xba}. We have shown that this result is, unusually, dependent on coupling the effective metric to a fundamental field, as the no-go theorem is specifically avoided because the pressure of such matter depends on the lapse function. This lapse dependence is not present in the perfect-fluid description typically employed in late-time cosmological setups, such as radiation ($p\sim a_\eff^{-4}$) and dust ($p=0$), and therefore a universe containing \emph{only} such matter will still run afoul of the no-go theorem. While this may not be a strong physical criterion --- cosmological matter is still built out of fundamental fields --- it presents a sharp practical problem in relating the theory to cosmological observations. If we assume that matter is described by perfectly pressureless dust, which is sensible on very large scales, then even the field description might not be sufficient, as the absence of pressure would set the right-hand side of \cref{eq:bianchi} to zero. Furthermore, if one uses a scalar field to avoid the no-go theorem, it cannot live on a flat potential and must be rolling. The latter consideration would seem to rule out the use of the Higgs field to unlock massive cosmologies, as we expect it to reside in its minimum cosmologically.

Overall, in principle one can obtain observationally-sensible cosmologies in doubly-coupled massive gravity, but either a new degree of freedom must be included, such as a new dark field or some other matter source with a nontrivial pressure, or we must treat cosmological matter in terms of their constituent fields. Thus we cannot apply the standard techniques of late-time cosmology to this theory.

We have further shown that if dust and radiation are doubly coupled as well --- which is necessary if we demand the new scalar matter obey the equivalence principle --- then the cosmologies generically are unable to reproduce a viable radiation-dominated era, and in the far future the Hubble rate diverges, rather than settling to a constant and producing a late-time accelerated expansion. These pathologies can only be avoided if the scalar field potential is highly contrived with tuned theory parameters, or dust and radiation do not doubly couple. In the latter case, there is generically late-time acceleration, but for much of the parameter space this is driven in large part by the potential of the scalar field. In those cases the modification to general relativity may not be especially well motivated by cosmological concerns, as the scalar field would play the role of dark energy and not provide much benefit over simple quintessence. Otherwise, the parameters of the theory need to be tuned to ensure that the theory truly self-accelerates.

It seems that the dRGT massive gravity only has viable FLRW cosmological solutions --- i.e., that evade the no-go theorems on existence \cite{D'Amico:2011jj} and stability \cite{DeFelice:2013awa} --- if one either includes a scalar field or some other ``exotic'' matter with a lapse-dependent pressure (or possibly a pressure depending on $\dot a$) and couples it to the effective metric proposed in \rcite{deRham:2014naa} or goes beyond the perfect-fluid description of matter. Even if one includes a new scalar degree of freedom, significant pathologies arise if normal matter couples to the same effective metric. In all setups, the need for descriptions beyond a simple perfect fluid makes this theory problematic from an observational standpoint. Indeed, one might compare this to the situation with the original dRGT theory, in which all matter couples to $g_\mn$. While FLRW solutions do not exist in this case, it is possible by mildly breaking the assumption of isotropy and homogeneity to evade the no-go theorem \cite{D'Amico:2011jj}. The real problem is that by dropping the highly-symmetric FLRW ansatz, we lose a great deal of predictability and it becomes significantly more difficult to unambiguously compare the theory to observations.

We end with three small caveats. Notice that we have assumed that in unitary gauge for the St\"uckelberg fields, i.e., choosing coordinates such that $\eta_\mn=\operatorname{diag}(-1,1,1,1)$, the metric has the usual FLRW form (\ref{eq:FRW}). However, that form is arrived at by taking coordinate transformations of a more general homogeneous and isotropic metric, so that assumption may be overly restrictive.\footnote{We thank Fawad Hassan for pointing this out to us.} Equivalently, one could consider a more general, inhomogeneous and/or anisotropic gauge for the St\"uckelbergs.

We also note that if this theory does possess a ghost, even with a mass above the strong-coupling scale, solutions to the nonlinear equations of motion could contain the ghost mode and therefore not be physical.\footnote{We thank Angnis Schmidt-May for discussions on this point.} In other words, the ghost-free effective theory below the strong coupling scale and the theory we have been studying may not have coinciding solutions. However, a Hamiltonian analysis showed that the ghost does not appear around FLRW backgrounds \cite{deRham:2014naa}, suggesting that we have studied the correct cosmological solutions to any underlying ghost-free theory.

Finally, if one simply gives dynamics to the reference metric, we end up with a theory of doubly-coupled bigravity which treats the two metrics on completely equal footing and has been shown to produce observationally viable cosmologies \cite{Enander:2014xga}, although some of the issues with doubly-coupled massive gravity, such as the potential ghost problem, will still remain.

\acknowledgments

We are grateful to Luca Amendola, Claudia de Rham, Matteo Fasiello, Fawad Hassan, Lavinia Heisenberg, Mikica Kocic, Johannes Noller, Raquel Ribeiro, Angnis Schmidt-May, and Andrew Tolley for highly enlightening discussions. A.R.S. has been supported by the David Gledhill Research Studentship, Sidney Sussex College, University of Cambridge; the Isaac Newton Fund and Studentships, University of Cambridge; the Cambridge Philosophical Society; and the STFC. Y.A. is supported by the European Research Council (ERC) Starting Grant StG2010-257080. Y.A. also acknowledges support from DFG through the project TRR33 ``The Dark Universe.'' F.K. acknowledges support from DFG Graduiertenkolleg GRK 1940 ``Particle Physics Beyond the Standard Model.'' E.M. acknowledges support for this study from the Swedish Research Council.

\bibliographystyle{JHEP}
\bibliography{bibliography}

\end{document}